\documentclass[prl,showpacs,superscriptaddress,twocolumn]{revtex4-1}
\usepackage{amssymb,amsmath,xcolor,graphicx,psfrag}

\usepackage{hyperref, cancel}

\begin{document}

%\nemmand{\jav}[1]{#1}
\newcommand{\jav}[1]{{\color{red}#1}}

\title{Non-Hermitian Lindhard function and Friedel oscillations}

\author{Bal\'azs D\'ora}
\email{dora.balazs@ttk.bme.hu}
\affiliation{MTA-BME Lend\"ulet Topology and Correlation Research Group,
Budapest University of Technology and Economics, 1521 Budapest, Hungary}
\affiliation{Department of Theoretical Physics, Budapest University of Technology and Economics, Budapest, Hungary}
\author{Doru Sticlet}
\email{doru.sticlet@itim-cj.ro}
\affiliation{National Institute for R\&D of Isotopic and Molecular Technologies, 67-103 Donat, 400293 Cluj-Napoca, Romania}
\author{C\u{a}t\u{a}lin Pa\c{s}cu Moca}
\email{mocap@uoradea.ro}
\affiliation{MTA-BME Quantum Dynamics and Correlations Research Group, Institute of Physics, Budapest University of Technology and Economics, Budafoki ut 8., H-1111 Budapest, Hungary}
\affiliation{Department  of  Physics,  University  of  Oradea,  410087,  Oradea,  Romania}
%\noaffiliation

\date{\today}

\begin{abstract}
The Lindhard function represents the basic building block of many-body physics and accounts for charge response, plasmons, screening, Friedel oscillation, RKKY interaction etc.
Here we study its non-Hermitian version in one dimension, where quantum effects are traditionally enhanced due to spatial confinement,
 and analyze its behavior in various limits of interest. Most importantly, we find that the static limit of the non-Hermitian Lindhard 
function has no divergence at twice the Fermi wavenumber and vanishes identically for all other wavenumbers at zero temperature.  
Consequently, no Friedel oscillations are induced by a non-Hermitian, imaginary impurity to lowest order in the impurity potential at zero temperature.
Our findings are corroborated numerically on a tight-binding ring by switching on a weak real or imaginary potential.
We identify conventional Friedel oscillations or heavily suppressed density response, respectively.

\end{abstract}

\maketitle

\paragraph{Introduction.}
Linear response theory is ubiquitous is various fields of physics~\cite{vignale}.  It is of vital importance to gain information 
on physical properties of matter and understand measurements in a lab.
One of the most prototypical examples involves the charge response of electrons, described by the Lindhard function. 
This sheds light on charge screening, optical properties, plasmon dispersion~\cite{Pines1952,hwang2007}, Friedel oscillations~\cite{Friedel1952,Friedel1958, Sprunger.1997, bena} 
and RKKY interaction~\cite{vignale}, to mention a few.
Besides its theoretical appeal of being the entry level many-body physics, 
the Lindhard function plays an important role in spin-glasses, spin transport~\cite{Vignale.2000}, 
plasmonics~\cite{Grigorenko2012}, and giant magnetoresistance~\cite{Stiles.1993,Fert2011}.

Friedel oscillations are of purely quantum mechanical origin~\cite{friedelgravity} and
manifest as spatially long-range rippling pattern of the particle density around an impurity or other defects.
Its spatial profile is closely related to the RKKY interaction between localized magnetic moments mediated by the itinerant electrons.
While Friedel oscillations are well documented in several systems~\cite{bena,dallatorre}, they have not been investigated  
in the context non-Hermitian quantum mechanics~\cite{El-Ganainy2019,Bergholtz2021,ashidareview}. 
This field of research already provided us with several unique phenomena such as unidirectional invisibility~\cite{Lin2011}, enhanced sensitivity~\cite{hodaei},
single-mode lasing~\cite{Feng2014} or exceptional points~\cite{heiss}.

It is therefore natural to ask what happens to the density profile around a non-Hermitian, purely imaginary impurity~\cite{Krapivsky.2014,Parris.1989}? 
Not only that such a scattering potential cannot be classified as repulsive or attractive, but also the adaptation of the Lindhard function to the non-Hermitian realm 
 is far from being obvious.

Using the recently developed non-Hermitian linear response theory~\cite{linresp1,linresp2,linresp3}, 
we study this problem and analyze the non-Hermitian Lindhard function, the response of free electrons to a non-Hermitian, imaginary potential scatterer. 
We focus on 1d at zero temperature where quantum effects are enhanced the most due to reduced dimensionality, 
as also testified by the conventional, Hermitian Lindhard function~\cite{mihaila}.
Its non-Hermitian counterpart displays completely different behavior,
and its imaginary part remains finite even outside of the electron-hole continuum~\cite{vignale}.
More importantly, the static non-Hermitian Lindhard function vanishes at zero temperature, indicating the \emph{absence} of Friedel oscillations in non-Hermitian systems. 
This conclusion remains valid also in 2d and 3d as well.
We test this numerically in a 1d tight-binding ring, after switching on a Hermitian or a non-Hermitian imaginary impurity. 
In both cases, light cones appear in the density profile, but only the former scenario contains conventional Friedel oscillations, while these are completely 
absent from the latter within the light cones, in accord with the non-Hermitian Lindhard function.

\paragraph{Free electrons in 1d with non-Hermitian perturbation.}
We consider a 1d free electron gas~\cite{Solyom2008,vignale} of spinless fermions,
\begin{gather}
H=\sum_k\epsilon(k)c^\dag_kc_k, \phantom{aaa} \epsilon(k)=\frac{k^2}{2m}
\end{gather}
in the presence of a local, manifestly non-Hermitian potential at the origin with strength $F$, which couples to the local density as
\begin{gather}
H'=iFn(0),\quad n(0)=n(R=0),
\end{gather}
where $n(R)=\frac 1L \sum_{k,q}\exp(iqR)c^\dag_{k+q}c_k$ is the particle density at position $R$ and $L$ is the system size. 
 The system 
is filled up to the chemical potential $\mu$, which defines the Fermi wavenumber $k_F$ as $\epsilon(k_F)=\mu$.
We are interested in the fate of the particle density variation after introducing $H'$ at $t=0$. 
%We consider a generic filling, up to a chemical potential $\mu$, which defines the Fermi wavenumber $k_F$ as $\epsilon(k_F)=\mu$.
%We are interested in the changes in the local density induced by the  non-Hermitian potential $H'$. 
To this end, we use the recently
developed non-Hermitian linear response theory~\cite{linresp1,linresp2,linresp3}.  
The associated response function describing the changes in $n(R)$ due to the imaginary, non-Hermitian potential  $H'$ is 
evaluated as
\begin{equation}
\chi(R,t)=-i\Theta(t)\left(\langle\left\{n(R),in(0)\right\}\rangle-2\langle n(R)\rangle\langle in(0)\rangle\right),
\label{chirt}
\end{equation}
where
%\begin{gather}
%n(R)=\frac 1L \sum_{k,q}\exp(iqR)c^\dag_{k+q}c_k,
%\end{gather}
$\{ , \}$ denotes the anticommutator and the expectation values are taken with respect to the thermal state of $H$.
Upon a Fourier transformation with respect to $R$ and $t$, and taking the expectation value using a standard procedure~\cite{Solyom2008,mahan},
we finally obtain the non-Hermitian Lindhard function as
\begin{gather}
\chi(q,\omega)=\frac{i}{L}\sum_{k}\frac{f(k)+f(k+q)-2f(k)f(k+q)}{\omega+i\eta+\epsilon(k)-\epsilon(k+q)},
\label{lindhard}
\end{gather}
where $\eta\rightarrow 0^+$, $f(k)=1/(\exp[(\epsilon(k)-\mu)/T]+1)$ is the Fermi function. 
We emphasize that this expression is valid for arbitrary dimensional electron gas.
While the denominator remains unchanged compared to the conventional Lindhard function~\cite{mihaila,Solyom2008,vignale,gruner}, 
there are still two notable differences: there is an extra $i$ prefactor and the numerator contains a different combination of the Fermi functions.
The latter stems from $f(k)(1-f(k+q))+f(k+q)(1-f(k))$, arising from
 the anticommutator in Eq.~\eqref{chirt}, while the Hermitian  case with the commutator~\cite{vignale} contains the difference
of these two terms.

\paragraph{Properties of the non-Hermitian Lindhard function.}
Let us start by investigating various limit. 
For our current discussion, the most important limit is the  static, $\omega=0$ limit. In this case, the denominator gives
$-i\pi\delta[\epsilon(k)-\epsilon(k+q)]$ in the $\eta\rightarrow 0^+$ limit, forcing energy conservation, and the non-Hermitian susceptibility becomes real  as
\begin{gather}
%\chi(q,0)=\frac{m}{\pi |q|}\left[f(q/2)-f^2(q/2)\right].
\chi(q,0)=\frac{m}{4\pi |q|}\cosh^{-2}\left(\frac{\epsilon(q/2)-\mu}{2T}\right)
\label{static}
\end{gather}
At finite temperatures, it displays a singularity~\footnote{This $1/|q|$ singularity is most probably a shortcoming of the linear response calculation,
higher order terms are expected to modify it to $1/(|q|+|q_0|)$ as in Ref.~\cite{froml}} for small $q$ as $\exp(-\mu/T)/|q|$, while it has a finite, 
non-diverging peak at $2k_F$ of $T$-independent size $m/(8\pi k_F)$ and width $\sim T$.
With decreasing temperature, it becomes non-analytic at $T=0$ in the sense that the Lindhard function is only non-zero  exactly at $q=2k_F$. 
This has measure zero in $k$-space and cannot contribute to the real space behavior of the Lindhard function after Fourier transform. 
This is to be contrasted to the log-divergent peak at $2k_F$ in the Hermitian case and the finite small momentum response.

The vanishing of the static non-Hermitian Lindhard function at $T=0$ occurs due to the peculiar conspiracy of the occupation number and energy conservation.
While the occupation number requires $k$ and $k+q$ states with opposite filling (i.e. empty or filled), similarly to the Hermitian case, 
the energy conservation requires their energies to be equal as $\epsilon(k)=\epsilon(k+q)$, unlike the Hermitian case.
These antithetical conditions cause the Lindhard function to vanish at zero temperature.

At $T=0$, the behavior of $\chi(q,0)$ indicates that there is practically no density response to an imaginary potential 
%in the thermodynamic limit
to linear order in the external potential. 
Consequently,  no conventional Friedel oscillations or RKKY interaction~\cite{vignale} are expected.
This also signals that the system has no weak coupling instability towards density wave order~\cite{giamarchi,gruner}.
We have also checked that the vanishing of the $T=0$ non-Hermitian static Lindhard function is not unique to 1d but happens also in 2d and 3d as well.
This implies absence of Friedel oscillations to leading order in the impurity potential also in 2d and 3d.

The other limit of immediate interest is the dynamical, $\omega\ne 0$ and $q=0$ limit. 
In this limit, the Hermitian response vanishes identically due to the fact that a time-dependent, spatially homogeneous perturbation, 
which couples to the total number of electrons, cannot induce any density response because the number of electrons is a conserved quantity.
In the present non-Hermitian setting, we get 
\begin{gather}
\chi(0,\omega)=\frac{2i}{\pi v_F}\frac{T}{\omega+i\eta}
\end{gather}
for the physically relevant, $T\ll\mu$ limit and $v_F=k_F/m$ is the Fermi velocity.
The response function vanishes for $T=0$ when the expectation values are taken with respect to the ground state and becomes finite only for a density matrix describing a thermal state. 

\begin{figure}[t]
\centering
\includegraphics[width=7cm]{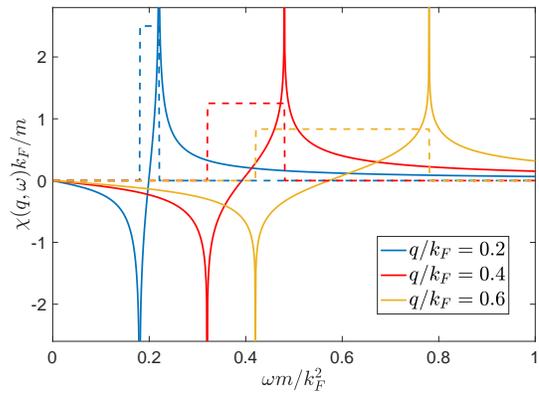}
\caption{The real (dashed lines) and imaginary (solid lines) part of the frequency dependent Lindhard function is plotted for several momenta at $T=0$. 
Compared to the Hermitian Lindhard function, the roles of the real and imaginary parts are reversed.}
\label{imchiT0}
\end{figure}
We also evaluate the real part of the non-Hermitian response function for arbitrary parameters as
\begin{gather}
\textmd{Re}\chi(q,\omega)=\frac{m}{2 |q|}\left[f(k_+)+f(k_-)-2f(k_+)f(k_-)\right],
\end{gather}
where $k_\pm=\frac{m}{q}\omega\pm\frac{q}{2}$. 
This reduces to Eq.~\eqref{static} in the static, $\omega=0$ limit.
As $T=0$, this expression is only non-zero for $k_+>k_F$ and $k_-<k_F$ for positive frequency and wavevector, and encodes information about the electron-hole continuum~\cite{vignale}. 
In this respect, it resembles to the imaginary part of the Hermitian Lindhard function, as shown in Fig.~\ref{imchiT0}, which is responsible for the structure factor.

At $T=0$, the imaginary part of the non-Hermitian Lindhard function is also determined analytically. 
In this limit, the Fermi functions reduce to Heaviside functions and the numerator of Eq.~\eqref{lindhard} is only finite if one of the Fermi functions is 1 while the other is 0. 
This gives the very same kinematic constraint as for the Hermitian case,
namely that $|k+q|$ and $k$ should be on opposite sides of $k_F$. 
However, the numerator of Eq.~\eqref{lindhard}, when non-zero, is 1, while it changes sign in the Hermitian case depending on whether $k$ is close to $\pm k_F$.
After some algebra, we obtain 
\begin{gather}
\textmd{Im}\chi(q,\omega)=-\frac{m}{2\pi |q|}\ln\left|
\dfrac{\left(\omega-q^2/2m\right)^2-v_F^2q^2}{\left(\omega+q^2/2m\right)^2-v_F^2q^2}\right|,
\label{imchi}
\end{gather}
which is an odd or even function of $\omega$ or $q$, respectively and its frequency dependence is depicted in Fig.~\ref{imchiT0}.
It diverges at $\omega_{\pm}=\mp q^2/2m\pm v_Fq\approx \pm v_Fq$ for small $q$ and positive frequency, which mark the boundary of the electron-hole continuum for the Hermitian 
case~\cite{vignale}.
However, $\textmd{Im}\chi(q,\omega)$ remains finite outside of $\omega_\pm$, and its behavior resembles to the real part of the Hermitian Lindhard function.

\paragraph{Tight-binding model with non-Hermitian impurity.}
The linear response theory describes the temporal evolution of an observable after an external perturbation was switched on. 
The static, $\omega=0$ limit of the Lindhard function predicts the long-time limit behavior of the local density after the perturbation was introduced.
Based on the vanishing static limit of the non-Hermitian Lindhard function at $T=0$, Friedel oscillation are not present around an 
imaginary impurity to lowest order in the potential strength~\cite{vignale}. 
In order to test the validity of this finding, we investigate numerically the Friedel oscillations in a 
half-filled 1d Hermitian tight-binding model with periodic boundary conditions (PBC), after switching on a local non-Hermitian impurity at $t=0$. 
In the long-time limit, the spatial profile is expected to be described by the static Lindhard function~\cite{vignale}.
We consider
\begin{gather}
H=\gamma\sum_{R=1}^N c^\dag_Rc_{R+1}+c^\dag_{R+1}c_R,
\label{hamtb}
\end{gather}
where the $c_R$'s are fermionic annihilation operators, $c_{N+1}=c_1$, and $\gamma$ is the hopping amplitude.
In order to avoid a degenerate Fermi gas~\cite{ehlers}, we choose $N=4m+2$ with integer $m$. 
The corresponding single particle spectrum is $\epsilon(k)=2\gamma\cos(k)$ with Fermi velocity $v_F = 2\gamma$.
The impurity is represented by
\begin{gather}
H'=Vc^\dag_{N/2}c_{N/2},
\label{imptb}
\end{gather}
which we consider either Hermitian for real $V$, or non-Hermitian for imaginary $V$.
We study $H+H'\Theta(t)$ numerically by solving the time-dependent Schr\"odinger equation. 
The initial many-body (i.e.~$N/2$-body) state, $\Psi_0$ is a Slater determinant made from the single particle eigenstates of Eq.~\eqref{hamtb}, 
denoted by $\phi_n$, satisfying $\langle \phi_n|\phi_n'\rangle=\delta_{n,n'}$.
Then, the time evolution is achieved at a single particle level as $\phi_n(t)=\exp[-i(H+H')t]\phi_n$ for $t>0$.
The time-evolved many-body wavefunction, $\Psi(t)$ remains a Slater determinant built up from these time-dependent single particle functions, which are no longer orthogonal to each,
since the time evolution is non-unitary due to the imaginary $V$ in $H'$, namely $\langle \phi_n(t)|\phi_n'(t)\rangle\neq \delta_{n,n'}$.
After time $t$, we evaluate numerically the change in the initially homogeneous density profile~\cite{carmichael,daley,ashidareview}
as
\begin{gather}
\delta n(R,t)=\frac{\langle\Psi(t)|c^\dag_Rc_R|\Psi(t)\rangle}{\langle\Psi(t)|\Psi(t)\rangle}-\frac 12,
\end{gather}
where the denominator in the first term is required as it accounts for the norm change of the wavefunction due to non-unitary dynamics~\cite{graefe2008}. 
Since the many-body wavefunction is a Slater determinant, $c_R$ in the numerator acts separately on the single particle wavefunctions. 
However, due to the non-orthogonality of $\phi_n(t)$, the overlap of the other wavefunctions, 
not acted on by $c_R$, has to be evaluated as well and can give non-trivial (i.e. not 0 or 1) contribution.

\begin{figure}[h!]
\centering
\includegraphics[width=8.5cm]{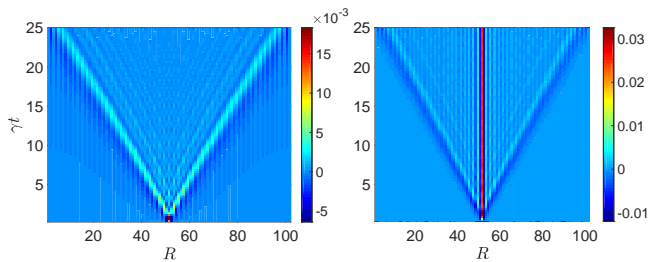}
\caption{Flat surface plot of the spatial and temporal evolution of the local 
density, for $N=102$ and $V=0.1i\gamma$ (non-Hermitian, left) and $V=-0.1\gamma$ (Hermitian, right) for Eq.~\eqref{imptb}. 
Inside the light cone, the density can readjust in response to the external potential, potentially causing Friedel oscillations. 
The distinct scales for the density variations indicate the suppression of Friedel oscillations in the non-Hermitian setting, compared to the Hermitian case.}
\label{friedel1}
\end{figure}

We have also solved the full many-body Hamiltonian using exact diagonalization, without making use of the Slater determinant nature of the wavefunction, for system sizes up to $N=26$. 
This approach agrees perfectly with the Slater determinant based single-particle numerics. 
However, the main advantage of time evolving the single particle states of $H$ and building up the many-body wavefunction from them is,
 that we easily reach systems of few 100 lattice sites, and therefore the time evolution is followed for sufficiently long times before finite size effects kick in.
Due to the pole structure of the non-Hermitian
Lindhard function from Eq.~\eqref{imchi} at $\omega\approx \pm v_Fq$, light-cone features are expected at $2\gamma t=\pm R$. 
For comparison, we also simulate numerically the case of a Hermitian impurity, when $H'$ with real $V$ is quenched suddenly at $t=0$.
These features are illustrated in Fig.~\eqref{friedel1}. For longer times, collapse and revival would take place, which are outside of the  scope of the present study.

\begin{figure}[h!]
\centering
\includegraphics[width=7cm]{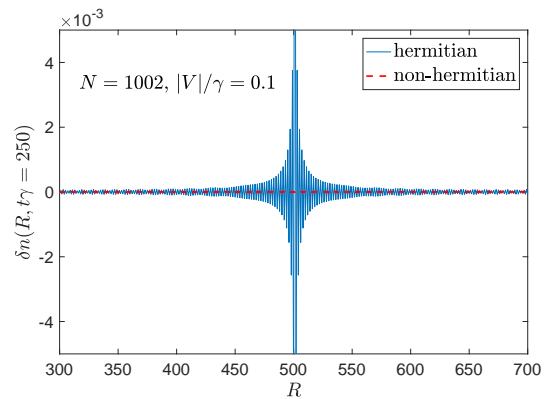}
\caption{Friedel oscillation of the half-filled tight-binding model with PBC and $N=1002$, the Hermitian or imaginary impurity potential is switched on at $t=0$ and
 $R=501$ with strength $V=0.1i \gamma$ (red, non-Hermitian) or $-0.1\gamma$ (blue, Hermitian), respectively.
The Friedel oscillations are negligible for an imaginary potential, in agreement with the prediction of the non-Hermitian Lindhard function from Eq.~\eqref{static} at $T=0$.}
\label{friedelmain}
\end{figure}

At time $2\gamma t=N/2$, we stop the time evolution before finite-size effects appear, namely when the two light cones meet at the opposite end of the tight-binding ring due to PBC.
For $N=1002$, this is shown in Fig.~\ref{friedelmain} for both the Hermitian and non-Hermitian cases. 
The Hermitian case follows the expected pattern of Friedel oscillation~\cite{vignale} with periodicity $1/2k_F$ and $|R-N/2|^{-1}$ decay of the envelope function. 
We have checked numerically that if we do \emph{not} quench the impurity potential, but determine the ground state of the 
Hermitian Hamiltonian $H+H'$ with real $V$ and evaluate the resulting time-independent density profile, it is indistinguishable 
from the previously evaluated $\delta n(R,t)$ in the long-time limit, $t=N/4\gamma$ after the quench.
In contrast, the non-Hermitian response is almost non-existent and hardly visible, since the first order contribution in the external potential vanishes in accord
with the prediction of the non-Hermitian Lindhard function,
and higher order terms are significantly suppressed due to the smallness of the potential, i.e. $|V|/\gamma\ll 1$.
We stress that this conclusion is true for spatio-temporal regions within the light cone, while the presence of the light cone in Fig.~\ref{friedel1} clearly manifests itself 
during the dynamics. This is readily expected from the pole structure of the Lindhard function from Fig.~\ref{imchiT0}.

Another way to address non-Hermitian Friedel oscillations invokes adiabatic continuity~\cite{pwanderson}. 
The ground state of Eq.~\eqref{hamtb} can be determined for any $N$. 
Then, by gradually introducing $V$, 
the evolution of the $V=0$ ground-state energy is followed in the complex energy plane by many-body diagonalizing $H+H'$ with a given 
imaginary $V$, and the corresponding wavefunction is also obtained.
Upon reaching the final imaginary value of $V$, this adiabatically continued wavefunction is used in determining the density pattern, which again gives no appreciable Friedel oscillations.

Finally, we also studied many-body interaction effects and found no visible Friedel oscillations for a non-Hermitian impurity, similarly to Fig. \ref{friedelmain}.
By adding the nearest neighbour interaction term $U\sum_{R=1}^Nc^\dag_Rc_{R}c^\dag_{R+1}c_{R+1}$ to $H$ in Eq. \eqref{hamtb} with $U=\gamma/2$, the Luttinger liquid ground state\cite{giamarchi} 
was determined using DMRG\cite{White-1992} with $N=100$
and the time evolution of the local densities after switching on $H'$ was monitored using TEBD\cite{Vidal-2007}. The numerics indicate that a Hermitian or non-Hermitian impurity 
induces conventional or negligible Friedel oscillation, respectively, similarly to the non-interacting case.

\paragraph{Experimental possibilities.}
Our non-Hermitian setting can originate from an open quantum system description based on the Lindblad equation. By introducing a coupling to the environment 
through localized particle gain or loss term~\cite{Syassen,Labouvie,Mullers,froml,Sels}, continuously monitoring the environment to make sure that no quantum jump~\cite{daley} 
occurs and postselecting the corresponding data yield exactly the type of non-Hermitian time evolution we consider.
Measuring the Friedel oscillations can be achieved using standard experimental tools. 
We note that in the full Lindblad description of our problem~\cite{froml}, conventional Friedel oscillations are present,
they are only suppressed when the time evolution is governed by a non-Hermitian Hamiltonian. It would be interesting to see 
how these features are re-established by departing from the non-Hermitian setting and allowing for quantum jumps.

\paragraph{Conclusion.}
We derived the non-Hermitian Lindhard function, the response function of a 1d electron gas to an imaginary external potential.
Its general frequency and momentum dependence differs from that in the Hermitian case. Its static limit vanishes for all wavenumbers at zero temperature.
This occurs because the usual requirement of filled initial and empty final state is unusually supplemented by energy conservation in the non-Hermitian case, 
requiring the initial and final states to possess the same energy, which is impossible to fulfill at zero temperature.
This leads to no Friedel oscillations around a non-Hermitian, imaginary impurity, which remains true in higher dimensions as well.
We benchmark our findings numerically on a tight-binding ring, which indicates the presence/absence of Friedel oscillations in the 
long-time limit after switching on a Hermitian/non-Hermitian impurity, respectively.
These features are measurable in a dissipative  open quantum system under continuous monitoring and postselection.

\begin{acknowledgments}
This research is supported by the National Research, Development and Innovation Office - NKFIH  within the Quantum Technology National Excellence Program (Project No.~2017-1.2.1-NKP-2017-00001), K119442, K134437, by the BME-Nanotechnology FIKP grant (BME FIKP-NAT), and by a grant of the Ministry of Research, Innovation and Digitization, CNCS/CCCDI–UEFISCDI, under projects number PN-III-P4-ID-PCE-2020-0277 and  PN-III-P1-1.1-TE-2019-0423, within PNCDI III.
\end{acknowledgments}

\bibliographystyle{apsrev}
\bibliography{wboson1}

\end{document}